\documentclass[11pt]{article}
\usepackage{geometry}                
\usepackage{wrapfig}
\usepackage{graphicx}
\usepackage{amssymb}
\usepackage{amsmath}
\usepackage{epstopdf}

\large\normalsize

\title{Collective behavior in coupled dynamical systems on two-dimensional weighted networks: A step toward understanding adaptive behavior of true slime mold
}
\author{Yuki Kagawa\thanks{(Corresponding author) 
Department of Electrical Engineering and Bioscience, Waseda University, Tokyo 169-8555, Japan 
\hspace{1mm} 
E-mail: \texttt{ykagawa@aoni.waseda.jp} 
\hspace{1mm} 
Present address: 
Institute for Nanoscience and Nanotechnology, Waseda University, Tokyo 162-8480, Japan
 } 
 and 
 Atsuko Takamatsu\thanks{
Department of Electrical Engineering and Bioscience, Waseda University, Tokyo 169-8555, Japan 
} 
 }

\date{}                                           

\begin{document}
\maketitle

\begin{abstract}
Plasmodium of true slime mold, \textit{Physarum polycephalum}, is an amoeboid organism, which spreads with developing tubular network structure and crawls on two-dimensional plane with oscillating the cell thickness. 
The plasmodium transforms its tubular network structure to adapt to the environment. 
To reveal the effect of the network structure on the oscillating behavior of the plasmodium, we constructed coupled map systems on two-dimensional weighted networks 
as models of the plasmodium, and investigated the relation between the distribution of weights on the network edges and the synchronization in the system. 
We found the probability that the system shows phase synchronization changes drastically with the weight distribution even if the total weight is constant. 
This implies the oscillating patterns observed in the plasmodium are controlled by the tube widths or cross-sections in the tubular networks. 
\end{abstract}

\section{Introduction}

\begin{figure}
\begin{center}
\includegraphics[width=8.6cm]{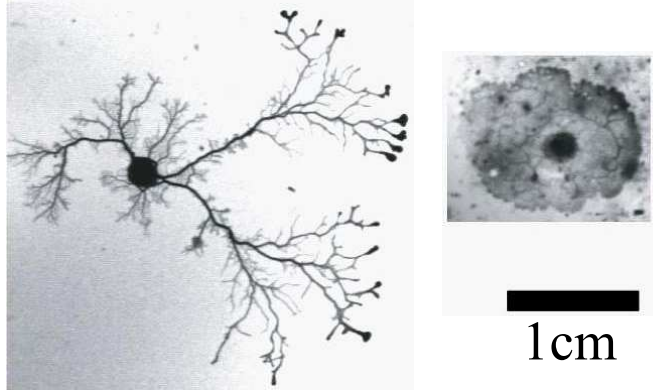}
\end{center}
\caption{Morphology of plasmodium of true slime mold. Tubular network structures observed on agar medium with 1 mM KCl (left), and oat flakes (right). Bar, 1cm. }
\label{slimenet}
\end{figure}

Plasmodium of true slime mold, \textit{Physarum polycephalum}, is a unicellular  amoeboid organism. 
It spreads with developing tubular network structure and crawls on two-dimensional plane. 
Partial bodies of the plasmodium, whose cell thickness oscillate in period 1-2 min, interact with each other through protoplasmic streaming in the tubes \cite{kamiya1981pac}. 
The network structure of the tubes, which determines the interactions between oscillating partial bodies, has environmental dependency \cite{takamatsu2009edm}. 
For example, the plasmodium shows dendritic structure when the culture medium contains harmful chemicals, whereas it shows disk structure when the medium contains oat flakes, food for this organism (see Fig.\ \ref{slimenet}). 
From the viewpoint of the network theory \cite{newman2006sad, boccaletti2006cns}, 
these dendritic and disk structures can be characterized as tree-like and lattice-like networks, respectively \cite{takamatsu2007nga}. 

Therefore, the plasmodium can be modeled as a two-dimensional network consisting of vertices that represent dynamical units, and edges that represent the interactions between the units \cite{kagawa2009pre}. 
In real plasmodium networks, widths of the tubes are not uniform but distributed largely. 
Because the width of the tube can be considered as the coupling strength in coupled biological oscillators of \textit{Physarum} plasmodium \cite{takamatsu2000cga, takamatsu2000tde}, each edge in a plasmodium network has its own "weight" that is directly connected to the coupling strength in the corresponding coupled dynamical systems. 
Thus, dynamical oscillation patterns observed in \textit{Physarum} are thought to be controlled not only by the structure of the network, but also by the weight distribution on a network. 

Recently we investigated the latter effect by varying the weighting rules on a fixed network topology; this topology was designed as an intermixture of tree and lattice so that we can express observed plasmodial tube networks \cite{kagawa2009pre}. 
On these generated two-dimensional weighted networks, coupled phase oscillators are constructed and 
the propensity for collective behavior, such as phase-locking and synchronous oscillation patterns, was investigated. 
As a result, we found that the distribution of edge weights in the networks strongly affects the global synchronization and spatiotemporal oscillation patterns, even if the network topology and the total weight are fixed. 

Can we apply this result to other coupled dynamical systems in general?
As a step toward answering this question, 
in the present paper, 
we construct coupled logistic maps, instead of phase oscillators, on two-dimensional weighted networks, and investigate their dynamics in detail. 

\section{Model}

\subsection{Weighted planer networks}

To express the environment-dependent plasmodial tube networks, 
two-dimensional weighted networks 
whose topology is an intermixture of tree and lattice 
are constructed as described in ref.\cite{kagawa2009pre}. 

\begin{figure}
\begin{center}
\includegraphics[width=4cm]{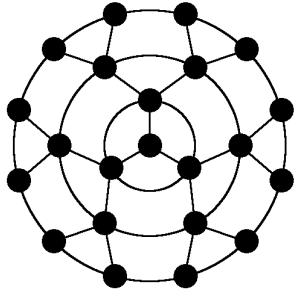}
\end{center}
\caption{Designed topology with $b=2$ and $d_{max}=3$. }
\label{topo}
\end{figure}

A network topology that includes a tree as well as a lattice is generated as follows. 
At first, a Cayley-tree with the branching number of $b$ and the maximum depth of $d_{max}$ is generated. 
The depth, $d$, of a given vertex is defined as the shortest path length from the tree's root to the vertex. Thus the depth of the root vertex is zero ($d=0$). 
The number of vertices in depth $d$ is given by $n(d) = (b+1)\cdot b^{d-1}$ when $d \ge 1$. 
Thus this tree has $N = 1 + \sum_{d=1}^{d_{max}} n(d) = 1+(b+1)(b^{d_{max}}-1)/(b-1)$ vertices, and $N-1$ edges. 
Note that the degree (connectivity) of non-boundary vertex (i.e. vertices in the depth $d \ne d_{max}$) is $b+1$ in this tree. 
Next, the vertices in the same depth are connected side by side to make a ring. 
In this process, $N-1$ edges are newly added, and all vertices but the root increase their degrees by two. 
The resulting topology with $b=2$ and $d_{max} = 3$ is shown in Fig.\ \ref{topo}. 

Each edge on the designed topology is weighted using simple symmetrical rules with a small number of parameters. 
Firstly, we separate the edges into two groups. 
The first group consists of $N-1$ edges originally belonging to the Cayley-tree, and the second group consists of the remainder $N-1$ edges newly added to the tree. 
Below, weights on the edges belonging to the first and the second groups are denoted by $w$ and $v$, respectively. 
Secondly, we assume the weights in a same depth are identical. 
Accordingly, the weights can be determined by a single parameter $d$, such that  $w(d)$ and $v(d)$.  
We define $w(d)$ as the weight on the edge that connects a vertex in depth $d-1$ and one in depth $d$. 
Similarly $v(d)$ is defined as the weight on the edge that connects vertices in the same depth $d$. 
Thirdly, we define the following parameters; $R=w(d+1)/w(d)$, $S=v(d)/w(d)$, and the total weights $w_{tot}$ given by 
\begin{equation}
w_{tot} = \sum_{d=1}^{d_{max}} n(d) [ w(d) + v(d) ]. 
\label{wtot}
\end{equation}

\begin{figure}
\begin{center}
\includegraphics*[width=12cm]{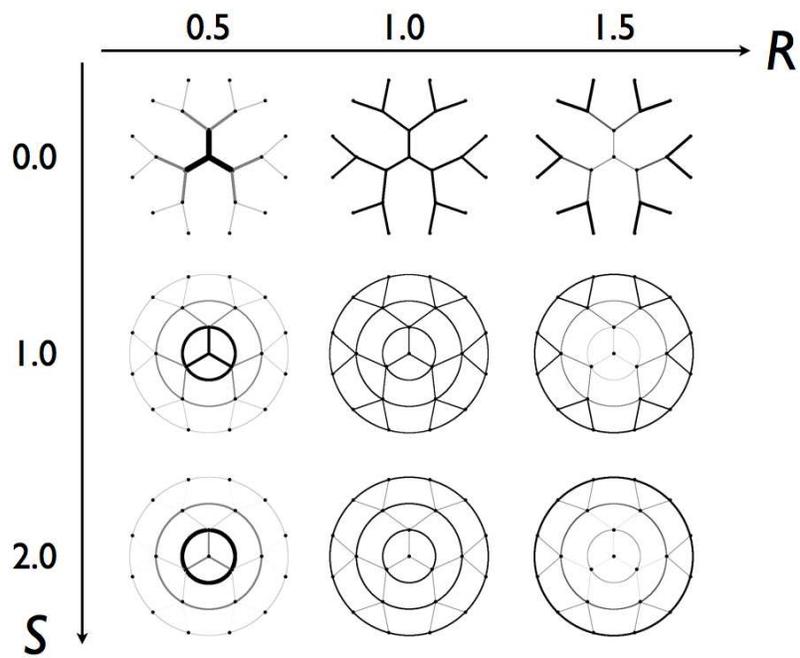}
\end{center}
\caption{Effect of the ratios $R$ and $S$ on the appearance of the weighted network. 
$b=2$ and $d_{max}=3$ are used. }
\label{RvsS}
\end{figure}

For given $R, S$, and $w_{tot}$, all of the weights on the edges in the network with the branching number $b$ and the maximum depth $d_{max}$ are uniquely determined as 
\begin{equation}
w(d) = 
	\begin{cases}
		\displaystyle\frac{ w_{tot} R^{d-1} (bR-1) }{(1+b)(1+S)[(bR)^{d_{max}}-1]}, & bR \ne 1, \\
		\displaystyle\frac{ w_{tot} R^{d-1}}{d_{max} (1+b)(1+S)}, & bR = 1, 
	\end{cases}
\label{wd}
\end{equation}
and $v(d) = S w(d)$, for $d=1, 2, \ldots, d_{max}$. 
In Fig.\ \ref{RvsS}, we illustrate how the ratios $R$ and $S$ affect the appearance of one of the designed networks when $w_{tot}$ is fixed. 
The parameter $R$ determines whether the network is center weighted or periphery weighted; the networks becomes center (periphery) weighted ones when $R<1$ ($R>1$). 
Another parameter $S$ determines whether the network is tree-like or ring-like; the networks becomes tree-like (ring-like) ones when $S<1$ ($S>1$).

\subsection{Coupled maps}
On the weighted networks characterized by $b, d_{max}, R, S$, and $w_{tot}$, coupled maps are constructed by putting a map on each vertex of the network and coupling the nearest neighbors. 
Dynamics of the coupled maps are given by 
\begin{equation}
x_i (t+1) = f_i[x_i(t)] + \sum_j A_{ij}  \{ f_j[x_j(t)] - f_i[x_i(t)] \}, 
\label{xit}
\end{equation}
where $x_i(t)$ is the state of the vertex $i (=1, 2, \ldots, N)$ at discrete time $t$, the function $f_i$ expresses a nonlinear map that generally depends on the index $i$ of the vertex, and the matrix component $A_{ij}$ expresses the coupling strength between the vertices $i$ and $j$. 
We assume the coupling strength is directly proportional to the weights on the corresponding edge. 
Here we set the proportional constant to be 1. 
Thus, if vertices $i$ and $j$ are connected, $A_{ij} (= A_{ji})$ is given by 
(i) $A_{ij} = w(d)$, when the corresponding edge $e$ is in the first group of edges 
and the depth of the deeper vertex ($i$ or $j$) is $d$, or 
(ii) $A_{ij} = v(d)$, when $e$ is in the second group and the depth of the both vertices is $d$.

In this paper, we used the logistic maps as dynamical units of the system; 
\begin{equation}
f_i(x) = a_i x (1-x), 
\end{equation}
where $a_i$ is a dynamical parameter that governs the uncoupled map dynamics. 
Since the amoeboid organism \textit{Physarum} shows basically periodic behavior, we mainly consider the case when all of the dynamical parameters $a_i (i=1, 2, \ldots, N)$ of the logistic maps are $2+\sqrt{6}/2 \simeq 3.22$, the center of the region $[3, 1+\sqrt{6}]$, where uncoupled maps show stable period-2 solution. 

\subsection{Phase synchronization}
To estimate the propensity for collective behavior, we calculate the probability that the system shows phase synchronization ($P_{sync}$), which is defined and calculated as follows. 
After weighting on each edge in the designed topology, we start calculation of  Eq.(\ref{xit}) with initial conditions, $x_i(0), i = 1, 2, \ldots, N$, given randomly from $[0, 1]$. 
After $\tau_f = 3000$ iterations, we obtain the next $\tau_o = 1000$ values for each $i$; i.e. $x_i(\tau_f + 1), x_i(\tau_f + 2), \ldots, x_i(\tau_f + \tau_o)$ for $i = 1, 2, \ldots, N$. 
Next, we calculate the phase distance between all pairs of vertices \cite{jalan2005scc}. 
The phase distance $d_{ij}$ between vertices $i$ and $j$ is defined as 
\begin{equation}
d_{ij} = 1 - \frac{\mu_{ij}}{\max (\mu_i, \mu_j)}, 
\end{equation}
where $\mu_i$ is the number of times the observed state values of vertex $i$, $x_i(t)$, show local minima during the time interval of $[\tau_f+1, \tau_f+\tau_o]$, and $\mu_{ij}$ is the number of times the minima of $x_i(t)$ and $x_j(t)$ match with each other. 
When $d_{ij} = 0$, we say vertices $i$ and $j$ are phase synchronized \cite{jalan2003soa}. 
If $d_{ij} = 0$ for all pairs of vertices in a cluster of vertices, this cluster is called a phase synchronized cluster. 
If the size of such a cluster is identical to $N$, we say the whole system is phase synchronized. 

For each initial condition, we obtain the size of the largest cluster existed in the system and calculate the probability that the size is equal to the size of the system $N$.  
This probability is identical with the probability $P_{sync}$ that the system shows phase synchronization. 

Because the coupling term in Eq.(\ref{xit}) does not normalized with the strength $s_i = \sum_{j=1}^N A_{ij}$, the logistic map may take an invalid value that is outside of $[0, 1]$. 
In this case, we stopped the calculation and started a new calculation with a new initial condition. 
We count these invalid cases as data showing no phase synchronization for the calculation of $P_{sync}$.

\section{Results}

\subsection{Coupled maps on weighted trees}
At first, we restrict our attention to the weighted trees, the case when there is no connection between vertices in the same depth, or $S=0$. 
Here we consider the coupled identical maps with $a_i = 2+\sqrt{6}/2$ (for all $i$) on a Cayley tree with $b=2$ and $d_{max}=3$. 
Since $S$ is fixed to be zero, only two parameters $R$ and $W$ determine the weight distribution on the tree. 
The mean weight $W$ is given by $W=w_{tot}/(N-1) = w_{tot}/21$, because the tree has $N = 22$ vertices and $N-1=21$ edges. 

\begin{figure}
\begin{center}
\includegraphics*[width=15cm]{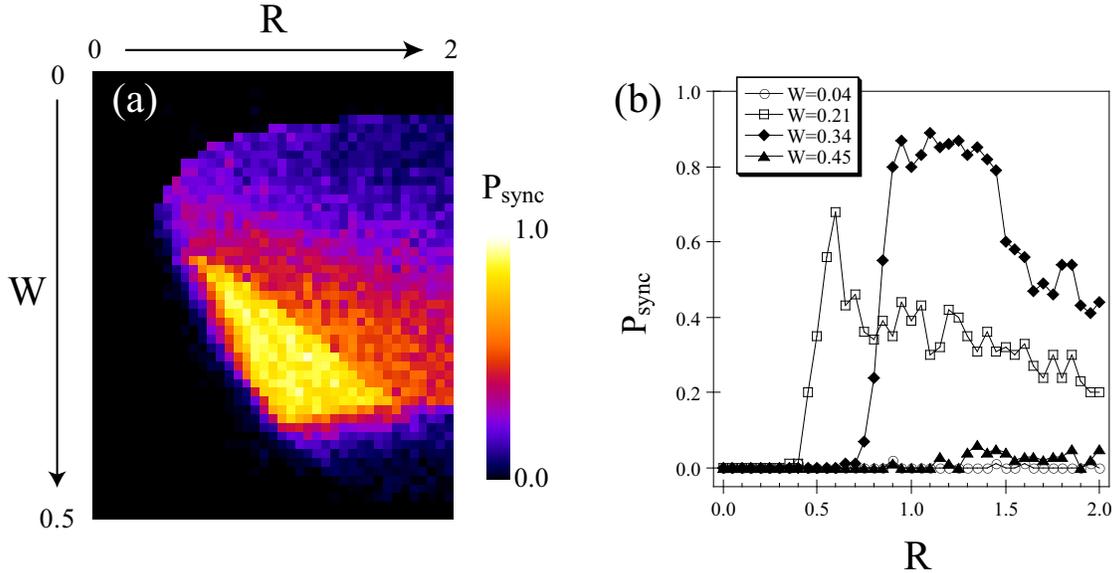}
\caption{
(a) Probability that the coupled maps on a weighted tree shows phase synchronization ($P_{sync}$) calculated in the $R$-$W$ space. 
Each probability for a given set of $R$ and $W$ is given by the number of samples showing the phase synchronization divided by the number of observations ($n=100$). 
(b) $P_{sync}$ plotted against $R$ for the cases of $W=$0.04 (open circles), 0.21 (open squares), 
0.34 (closed diamonds), and 0.45 (closed triangles). 
}
\label{tree_Psync}
\end{center}
\end{figure}

We calculate the probability $P_{sync}$ that the system shows phase synchronization in the space 
of $R$ and $W$ [Fig.\ \ref{tree_Psync}(a)]. 
This figure shows the dependence of $P_{sync}$ on weight distribution ($R$) strongly depends on the mean weight ($W$). 
In cases of small mean weights [e.g. $W=0.04$ shown in Fig.\ \ref{tree_Psync}(b)], the coupled maps rarely synchronize because the coupling strengths are too small. 
In cases of moderate mean weights, $P_{sync}$ has a value larger than 0.5 in the appropriate region of $R$. 
When $W$ is relatively small, e.g. $W=0.21$ shown in Fig.\ \ref{tree_Psync}(b), $P_{sync}$ has its largest value at the appropriate $R$ less than 1. 
This implies that the coupled map system on a center-weighted network ($R<1$) can phase synchronize easier than uniformly weighted ($R=1$) or out-of-center weighted ($R>1$) trees. 
On the other hand, when $W$ is relatively large, e.g. $W=0.34$ shown in Fig.\ \ref{tree_Psync}(b), $P_{sync}$ has its largest value at the appropriate $R$ between 1 and 1.5. 
This implies that uniformly or slightly out-of-center weighted trees are appropriate for the synchronization of the coupled maps. 
In cases of large mean weights, e.g. when $W=0.45$ shown in Fig.\ \ref{tree_Psync}(b), $P_{sync}$ has very small value regardless of $R$ because the maps frequently take invalid values outside of $[0, 1]$. 
With the same reason, $P_{sync}$ is almost zero in the region near the lower left corner of Fig.\ \ref{tree_Psync}(a), where $W$ is large and $R$ is small. 

We also calculate $P_{sync}$ in cases of randomly weighted trees, where each edge on a tree is weighted with a value given randomly from $[W-\Delta, W+\Delta]$ (data not shown).  
As a result, we found that the more randomly weighted on the edges of a tree, the less probable the coupled map system on the weighted tree is phase synchronized. 
This implies that the case of $\Delta = 0$, corresponding to a uniformly weighted tree, has $P_{sync}$ larger than any other randomly weighted cases at a given $W$. 
Note, however, that even in the case of $\Delta = 0$, $P_{sync}$ is smaller than the largest $P_{sync}$ obtained in the cases weighted orderly with the appropriate parameter $R$. 

\subsection{Coupled maps on the designed networks}
\begin{figure}
\begin{center}
\includegraphics*[width=12cm]{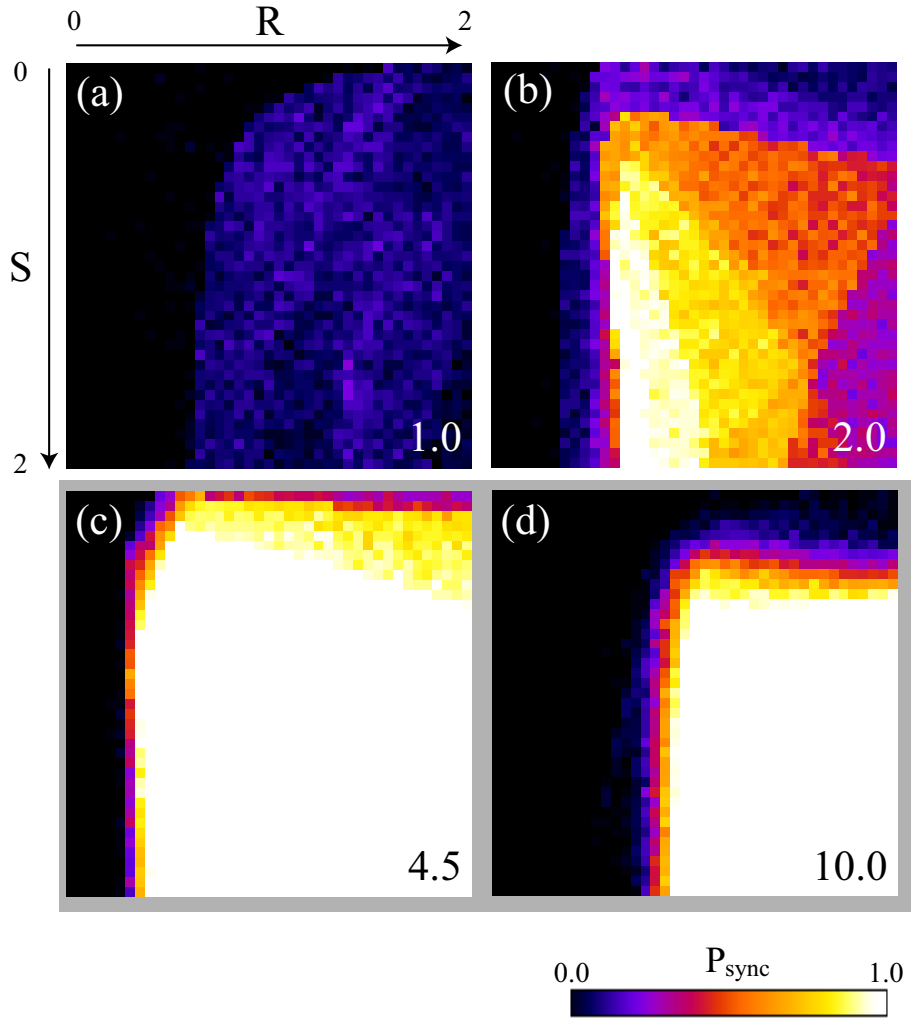}
\caption{
Probability that the coupled maps on the two-dimensional weighted designed network show the phase synchronization ($P_{sync}$) calculated in the $R$-$S$ space when the identical dynamical parameter $a_i = 2+\sqrt{6}/2$ (for all $i$)  is used for each logistic map. 
The total weight ($w_{tot}$) used for the calculation is shown in lower right corner in each figure. 
The number of observation for each parameter set of $R$ and $S$ is 100. 
$b=2$ and $d_{max}=3$ are used. 
}
\label{Psync}
\end{center}
\end{figure}

In the cases when there are connections between vertices in the same depth, or the cases of $S \ne 0$, 
we consider the coupled identical maps with $a_i = 2+\sqrt{6}/2 \hspace{3mm}$ (for all $i$) on the designed network with $b = 2$ and $d_{max}=3$. 
The probability that the system shows phase synchronization, $P_{sync}$, is calculated in the parameter space of $R$ and $S$ for a given total weight $w_{tot}$ (see Fig.\ \ref{Psync}). 
When $w_{tot}$ is small, for example when $w_{tot} = 1.0$, the system rarely phase synchronizes regardless of the parameters [see Fig.\ \ref{Psync}(a)]. 
On the other hand, when $w_{tot}$ is large, for example when $w_{tot} = 4.5$ or $10$, the system always phase synchronizes regardless of the parameters, as long as  the logistic maps take valid values, i.e. $[0, 1]$ [see Figs.\ \ref{Psync}(c) and (d)]. 
When the total weight is moderate, for example, in the case of $w_{tot} = 2.0$, $P_{sync}$ depends on the weight distribution determined by $R$ and $S$ [see Fig.\ \ref{Psync}(b)]. 
As shown in Fig.\ \ref{Psync}(b), $P_{sync}$ becomes very large when $S \ge 0.5$ and $0.6 \le R \le 1.0$, implying that the coupled maps on center-weighted ($R<1.0$) lattice-like ($S \ne 0$) networks are easy to phase-synchronize. 

\begin{figure}
\begin{center}
\includegraphics*[width=12cm]{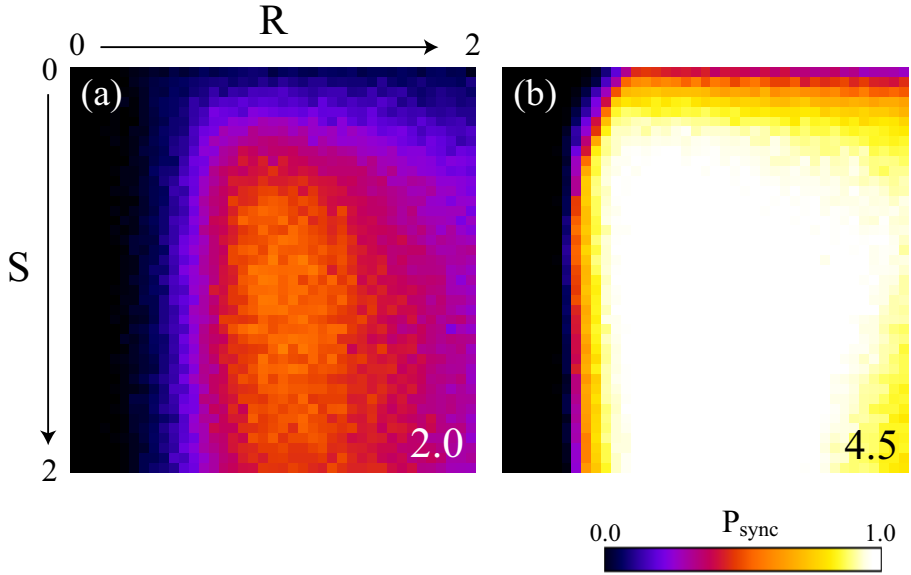}
\caption{ 
[(a) and (b)] Probability that the coupled maps on the two-dimensional weighted designed network ($P_{sync}$) calculated in the $R$-$S$ space when the dynamical parameter $a_i$ for each map is given randomly from the region $[1.8+\sqrt{6}/2, 2.2+\sqrt{6}/2]$ . 
The total weight ($w_{tot}$) used for the calculation is shown in lower right corner in each figure. 
Other parameters and calculation conditions are the same shown in Fig.\ \ref{Psync}. 
}
\label{Psync_randA}
\end{center}
\end{figure}

If the dynamical parameter $a_i$ of maps are not identical but varied uniformly between $1.8 + \sqrt{6}/2  \simeq 3.02$ and $2.2 + \sqrt{6}/2  \simeq 3.42$, the effect of weight distribution on the propensity for phase synchronization changes. 
Figures\ \ref{Psync_randA}(a) and (b) show the probabilities that such coupled maps show phase synchronization ($P_{sync}$) in the $R$-$S$ plane when the total weight is 2.0 and 4.5, respectively. 
In the case of $w_{tot} = 2.0$, the probability obtained in the system composed of non-identical maps has its peak at $R \sim 1$ and $S \sim 1$, corresponding to the uniformly weighted network, 
although the probability is smaller than that obtained in the system composed of identical maps [Fig\ \ref{Psync}(b)] for any given parameter set of $R$ and $S$. 

\section{Sufficient conditions that the system takes valid values}
As mentioned above, when $w_{tot}$ is large enough, the system always phase synchronizes as long as  the maps take valid values, i.e. $[0, 1]$ for the logistic map. 
In those cases, the region of the weight distribution $(R, S)$, where the coupled map systems defined by Eq.(\ref{xit}) are phase synchronized, can be derived by the sufficient conditions for the maps to take valid values. 
Such conditions are expressed mathematically as $\max [x_i(t+1)] \le 1$ (for all $i$), where $\max[\cdot]$ represents the maximum value. 

Since Eq.(\ref{xit}) can be rewritten as 
\begin{equation}
x_i(t+1) = (1-s_i) f_i + \sum_j A_{ij} f_j, 
\label{xit2}
\end{equation}
where $s_i = \sum_j A_{ij}$ is the strength of vertex $i$, we have 
\begin{equation}
\max[ x_i(t+1) ]  = 
\begin{cases}
(1-s_i) f_i^{max} + \sum_j A_{ij} f_j^{max}, & s_i \le 1, \\
(1-s_i) f_i^{min} + \sum_j A_{ij} f_j^{max},  & s_i > 1,
\end{cases} 
\label{max_xit}
\end{equation}
where we define the range of the function $f_i$ as $[f_i^{min}, f_i^{max}]$. 
When the ranges of all of the maps are identical to $[f^{min}, f^{max}]$, we have 
\begin{equation}
\max[ x_i(t+1) ]  = 
\begin{cases}
f^{max}, & s_i \le 1,\\
(1-s_i) f^{min} +s_i f^{max}, & s_i >1. 
\end{cases}
\end{equation}

In the cases using the logistic maps, we have $f^{min} = 0$ and $f^{max} = f(1/2) = a/4$, where $a$ is the dynamical parameter. 
While $\max[x_i(t+1)] = f^{max} = a/4$ is less than unity for any given $a(\le 4)$ when $s_i \le 1$, $\max[x_i(t+1)]$ will be less than unity if 
\begin{equation}
s_i \le \frac{1-f^{min}}{f^{max}-f^{min}} = \frac{4}{a}
\end{equation}
is satisfied when $s_i > 1$. 
Therefore, the sufficient conditions that each map in the coupled identical logistic maps with dynamical parameter $a$ takes values within the domain $[0,1]$ are 
\begin{equation}
\frac{1}{4} a s_i \le 1, \hspace{5mm} 
i = 1, 2, \ldots, N. 
\label{condition}
\end{equation}

If we use the weighting rules mentioned above, the strengths are uniquely determined by the depth of the vertex, $d$. 
Thus, the strength can be written as a function of $d$, i.e. $s(d)$. 
This function is given by 
\begin{equation}
s(d) = 
\begin{cases}
(b+1) w(1), & d=0,\\
(1+2S+bR) w(d), & d=1, 2, \ldots, d_{max}-1, \\
(1+2S) w(d_{max}), & d = d_{max}.
\end{cases}
\label{sd}
\end{equation}
In those equation, $w(d), d=1, 2, \ldots, d_{max}$ are given as Eq.(\ref{wd}). 
Using this function, the sufficient conditions can be rewritten as 
\begin{equation}
\frac{1}{4} a s(d) \le 1, \hspace{5mm} d = 0, 1, \ldots, d_{max}
\label{condition2}
\end{equation}

\subsection{The case of weighted trees}
In the case of the weighted tree (i.e. the case of $S=0$), the largest strength in the vertices is $s(0)$  when $R\le1$, and $s(d_{max}-1)$ when $R>1$. 
Thus the conditions (\ref{condition2}) are respectively given by $a s(0)/4 \le 1$ and $a s(d_{max}-1)/4 \le 1$. 
Substituting Eq.(\ref{sd}) into these equations, we have the sufficient conditions as follows. 
\begin{eqnarray}
W &\le& \frac{4(b-1)[(bR)^{d_{max}}-1] }{ a(b+1)(bR-1)(b^{d_{max}}-1)}, \hspace{5mm} {\rm when} \hspace{3mm} R\le1 \label{condition_tree1} \\
W &\le& \frac{4(b-1)[(bR)^{d_{max}}-1] }{ a R^{d_{max}-2} [(bR)^2-1] (b^{d_{max}}-1)},  \hspace{5mm} {\rm when} \hspace{3mm} R>1. 
 \label{condition_tree2} 
 \end{eqnarray}
In the special case of $bR=1$, where $R \le 1$ is satisfied because $b$ is an integer larger than 1, 
the condition is given by  
\begin{equation}
W \le \frac{4 d_{max} (b-1)}{a (b+1)(b^{d_{max}}-1)}. 
\end{equation}

\begin{figure}
\begin{center}
\includegraphics*[width=15cm]{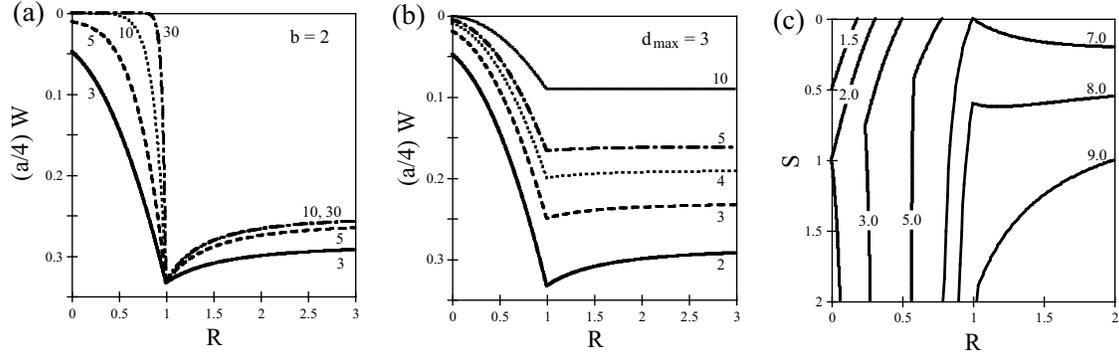}
\caption{ 
(a) The region in which all of the logistic maps on the weighted tree never take invalid values when $b=2$ and $d_{max} =$ 3, 5, 10, and 30 [Eqs.\ (\ref{condition_tree1}) and (\ref{condition_tree2})]. 
Each region includes the upper-right corner. 
(b) The same region when $d_{max}=3$ and $b=$2, 3, 4, 5, and 10. 
Each region includes the upper-right corner. 
(c) The region in which all of the maps on the weighted designed network never take invalid values 
when $b=2$ and $d_{max}=3$ [Eqs.\ (\ref{condition_1})-(\ref{condition_4})]. 
Each region includes the lower-right corner. 
Numbers on/near the solid lines represent $(a/4) w_{tot}$. 
}
\label{ValidRegion}
\end{center}
\end{figure}

Figure\ \ref{ValidRegion}(a) shows the region in which all of the maps never take invalid values when $b=2$ and $d_{max} =$ 3, 5, 10, and 30. 
The regions include the upper-right corner ($R=3$, $W=0$) of the figure. 
As the maximum depth increases, the region shrinks to that in the limiting condition of $d_{max} \to \infty$;
\begin{eqnarray}
W &=& 0, \hspace{5mm} {\rm when} \hspace{3mm} R\le1 \\
W &\le& \frac{4R^2(b-1)}{ a [(bR)^2-1]},  \hspace{5mm} {\rm when} \hspace{3mm} R>1. 
\end{eqnarray}

Figure\ \ref{ValidRegion}(b) shows the same region when $d_{max}=3$ and $b=$2, 3, 4, 5, and 10. 
The regions include the upper-right corner ($R=3$, $W=0$) of the figure. 
As the branching number increases, the region also shrinks to the upper.

\subsection{The case of weighted designed networks}
In the case of $S\ne0$, the sufficient conditions that each map in the coupled logistic maps takes values within the domain $[0,1]$ [i.e. Eq.(\ref{condition})] are given by 
\begin{eqnarray}
\frac{1}{4} a s(0) &\le& 1, \hspace{5mm} \textrm{when}\hspace{3mm} R \le 1-2S/b \label{condition_1}\\
\frac{1}{4} a s(1) &\le& 1, \hspace{5mm} \textrm{when}\hspace{3mm} 1-2S/b < R \le 1\\
\frac{1}{4} a s(d_{max}-1) &\le& 1, \hspace{5mm} \textrm{when}\hspace{3mm} 1 < R \le (1+2S)/(1+2S-b) \\
\frac{1}{4} a s(d_{max}) &\le& 1, \hspace{5mm} \textrm{when}\hspace{3mm} R > \frac{1+2S}{1+2S-b} (>1)
 \label{condition_4}
\end{eqnarray}
where $s(d), d=0, 1, d_{max}-1$, and $d_{max}$ are given by Eq.(\ref{sd}). 
Figure\ \ref{ValidRegion}(c) shows the boundaries of the region in which all of the maps never take invalid values when $b=2$ and $d_{max} = 3$. 
The region includes the lower-right corner ($R=S=2$). 
Numbers on or near the boundaries represent $(a/4)w_{tot}$. 
These boundaries can predict the region in which the system shows phase synchronization with high probability when $w_{tot}$ is large enough. 
For example, the region of $P_{sync}=1.0$ in the case of $w_{tot}=10$ [Fig.\ \ref{Psync}(D)] is similar to the region in which Eq.(\ref{condition2}) is satisfied when $(a/4)w_{tot} = [(2+\sqrt{6}/2)/4] \cdot 10 \simeq 8$ [see Fig.\ \ref{ValidRegion}(c)]. 

In the case of $w_{tot}=2.0$ and $a=2+\sqrt{6}/2$, used in the calculation shown in Fig.\ \ref{Psync}(b), 
the line of $(a/4)w_{tot} \simeq 1.6$ corresponds to the boundary. 
Thus the all region without the small region including the upper-left corner ($R=S=0$) satisfies the sufficient condition given by Eq.(\ref{condition2}) [see Fig.\ \ref{ValidRegion}(c)]. 
Therefore the dependency of the weight distribution on the probability $P_{sync}$ shown in Fig.\ \ref{Psync}(b) does not include the effect that the maps take invalid values.

\section{Discussion}

For many years, synchronization in coupled dynamical systems on complex networks has been studied by many researchers (see ref.\cite{osipov2007son} and references therein). 
Many of the studies considered a system composed of identical dynamical units on unweighted networks \cite{barahona2002ssw, nishikawa2003hon, moreno2004sko}. 
To elucidate synchronous phenomena observed in real networks \cite{barrat2004acw}, studies on synchronization of the systems on weighted networks are needed. 
Previous studies on coupled dynamical systems on weighted networks are based on the assumption that vertices in the networks have large degrees \cite{zhou2006usw}. 
However, degrees of the vertices in many real networks, including tubular networks developed by the plasmodium (Fig.\ \ref{slimenet}), are not so large. 
Therefore, in the present study, we have constructed coupled logistic maps 
on two-dimensional weighted networks whose vertices have small degrees ($ \le b+3$). 
Using this system, we have shown that the probability that the system shows phase synchronization depends strongly on the weight distribution when the total weight is moderate [see Fig.\ \ref{Psync}(b)]. 


We have also studied dynamics of phase oscillators on two-dimensional weighted networks \cite{kagawa2009pre}, whose topology and weighting rules are the same shown in the present study. 
The probability that the system shows phase locking was calculated in the parameter space of $R$ and $S$ using a moderate total weight $w_{tot}$. 
As a result, it is found that the high probability region of phase locking (Fig. 4 in ref. \cite{kagawa2009pre}) resembles the one of phase synchronization when the dynamical parameter $a_i$ for each map is randomly given (Fig.\ \ref{Psync_randA}). 
Therefore we speculate that the relation between probability showing collective behavior and local coupling strength, or weight distribution in a given network, is less dependent on types of dynamical systems, i.e., whether the system is composed of phase oscillators or maps. 


In conclusion, we studied the dynamics of coupled logistic maps on two-dimensional weighted networks
as models for plasmodium of true slime mold, \textit{Physarum polycephalum}, spreading with developing tubular networks. 
We found the weight distribution among edges of the networks strongly affects whether the coupled maps are phase synchronized even if the total weight is constant. 
This implies that synchronization in \textit{Physarum} networks is controlled by changes in the width, or the cross sectional area, of each plasmodial tube that determines the coupling strength between partial bodies of the system. 

\small  
\vspace*{1cm}
\subsubsection*{Acknowledgments}
This study was supported by a Grant-in-Aid for Scientific Research on Priority Areas 
"Emergence of Adaptive Motor Function through Interaction between Body, Brain and Environment" 
from the Japanese Ministry of Education, Culture, Sports, Science and Technology to A. T. 
\normalsize 

\bibliographystyle{unsrt} 

\end{document}